\documentclass[aps,pra,reprint,a4paper,showpacs]{revtex4-1}
\usepackage{amsmath}
\usepackage[latin9]{inputenc}
\usepackage{graphicx}
\usepackage{hyperref}
\usepackage{siunitx}
\usepackage{epstopdf}

\begin{document}

\title{Detrimental adsorbate fields in experiments with cold Rydberg gases near
surfaces}

\author{H. Hattermann}
\email{hattermann@pit.physik.uni-tuebingen.de}
\author{M. Mack}
\author{F. Karlewski}
\author{F. Jessen}
\author{D. Cano}
\author{J. Fort\'{a}gh}
\email{fortagh@uni-tuebingen.de}

\affiliation{CQ Center for Collective Quantum Phenomena and their
Applications, Physikalisches Institut, Eberhard-Karls-Universit\"at
T\"ubingen, Auf der Morgenstelle 14, D-72076 T\"ubingen,
Germany}

\begin{abstract}
We observe the shift of Rydberg levels of rubidium close to a copper surface when
atomic clouds are repeatedly deposited on it. We measure transition frequencies of
rubidium to $S$ and $D$ Rydberg states with principal quantum numbers $n$ between 31
and 48 using the technique of electromagnetically induced transparency. The
spectroscopic measurement shows a strong increase of electric fields towards the
surface that evolves with the deposition of atoms. Starting with a clean surface, we
measure the evolution of electrostatic fields in the range between 30 and
\SI{300}{\micro \meter} from the surface. We find that after the deposition of a
few hundred atomic clouds, each containing $\sim10^6$ atoms, the field of adsorbates
reaches \SI{1}{\volt/\centi\meter} for a distance of \SI{30}{\micro\meter} from the
surface. This evolution of the electrostatic field sets serious limitations on
cavity QED experiments proposed for Rydberg atoms on atom chips.  

\end{abstract}


\maketitle

\section{Rydberg atoms at surfaces}

The large electric polarizability of Rydberg atoms leads to a large response to
electric fields \cite{Gallagher1994}. This property is an enormous advantage for
applications that require fast coupling between atoms and photons, such as the
entanglement of Rydberg atoms via the electromagnetic modes of radio-frequency and
microwave cavities \cite{Raimond2001}. Several quantum computation schemes have been
proposed based on Rydberg atoms coupled to superconducting coplanar cavities
\cite{Petrosyan2008, Petrosyan2009, Saffman2010}. In the proposed scenarios, cold atomic gases are first positioned near a coplanar resonator. 
The atoms are subsequently laser excited into Rydberg states which interact with the electromagnetic modes of the resonator.
Recent progress with coupling such
cavities to superconducting qubits \cite{DiCarlo2009, Wallraff2004} and the coupling
of Rydberg atoms to a microwave stripline \cite{Hogan2012} outline good perspectives.

However, the technical realization faces challenges. One significant problem is the
detrimental effect of the electrostatic fields generated by adsorbed atoms on the
chip surface. Because of the high electronegativity of metals, atoms deposited on
the chip surface partially donate their valence electron to the metal. The result is
a permanent electric dipole layer on the surface that produces inhomogeneous
electrostatic fields and alters both the energy and orbital structure of nearby
Rydberg atoms. The fields can be strong enough to shift Rydberg states out of the
cavity resonance.

New chips are initially free of adsorbates, but experiments progressively accumulate adatoms on
the surface. An important question is how long it takes until the accumulation
of atoms on the surface becomes detrimental. Two research groups reported previously on this subject.
First, McGuirk et al. and Obrecht et al. studied the electrostatic field of adsorbed
atoms on both conducting and insulating surfaces \cite{McGuirk2004, Obrecht2007}.
Second, Tauschinsky et al. measured electrostatic fields of adsorbates using
electromagnetically induced transparency (EIT) on Rydberg states
\cite{Tauschinsky2010}. The results presented in this article complement the data
published by these two groups. We measure the evolution of electrostatic fields at
distances of 30 - \SI{300}{\micro\meter} to the surface during a series of
consecutive experiments. Starting with a clean copper surface we deposit clouds of
$^{87}$Rb atoms onto the surface and measure the inhomogeneous electrostatic field
of polarized adatoms by spectroscopy on Rydberg states. We find that the
electrostatic fields are already significant after about a few hundred experimental
cycles. This corresponds to only a few hours of operation for a typical cold atom
experiment.

\section{Measurement of the electrostatic fields of adsorbates by Rydberg EIT}

\begin{figure}
\centerline{\includegraphics[trim = 0 0 0 0, width=.5\textwidth]{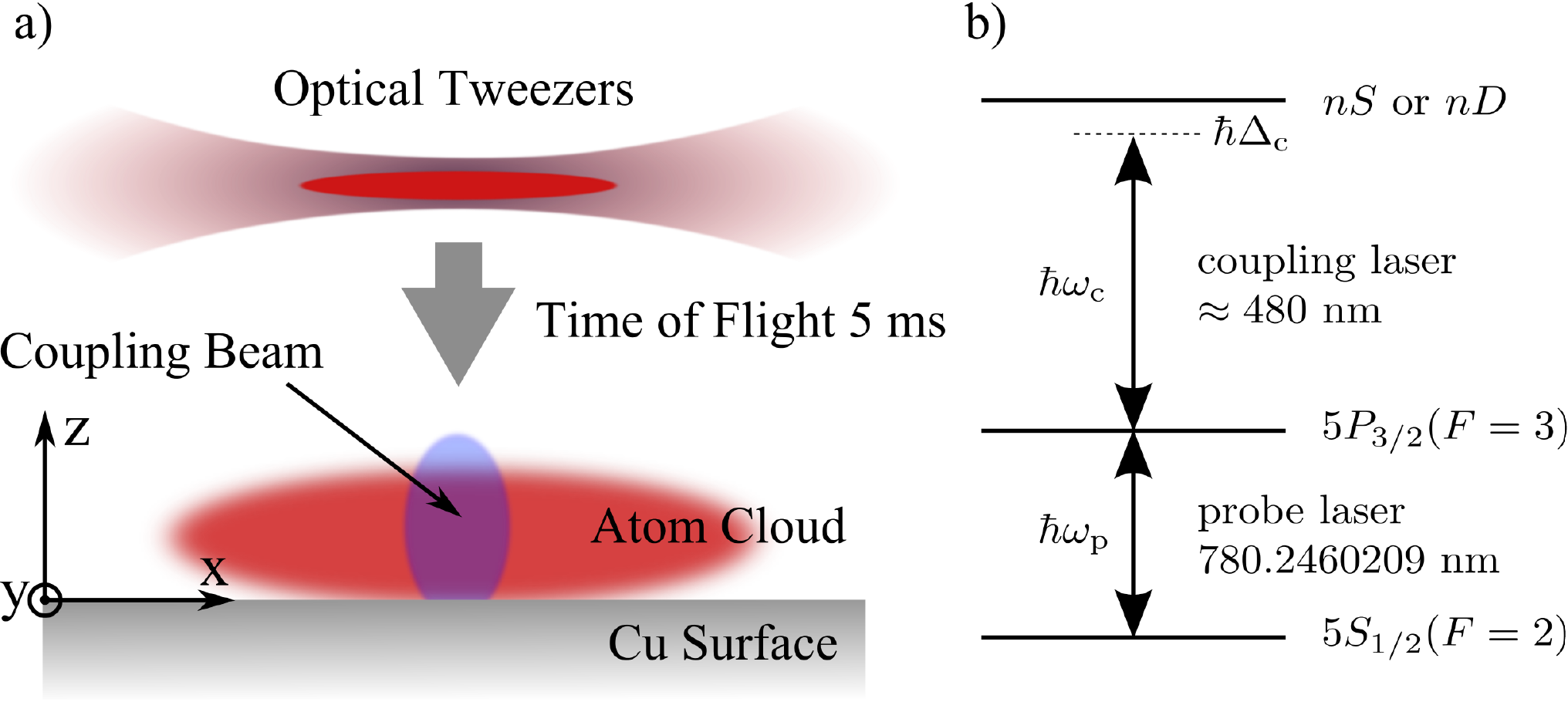}}
\caption{(a) An atomic cloud ($10^6$ $^{87}$Rb atoms in the $5S_{1/2} (F=2, m_F=2)$
ground state at temperature \SI{1.5}{\micro\kelvin}) is released from an optical
dipole trap to free fall and expansion. The cloud hits a Cu surface
\SI{200}{\micro\meter} below the trap. The electric field of polarized adatoms is
tested by means of Rydberg EIT. While the probe beam illuminates the full area, the
coupling beam has a smaller, elliptical spot size, illustrated in blue. (b)
Ladder-configuration for EIT signals. The probe beam couples the $5S_{1/2} (F=2,
m_F=2)$ and the $5P_{3/2} (F=3, m_F=3)$ levels. Simultaneously, a counter-propagating
coupling beam focussed onto a part of the cloud  (blue spot) is switched on. The
lasers are referenced to a frequency comb. In Ref. \cite{Mack11} we describe
the details of the locking method.}
\label{scheme}
\end{figure}

We measure the electrostatic field of adsorbed and polarized adatoms through the
energy shift (DC Stark shift) induced on highly excited Rydberg states of rubidium.
We start our experiments with a clean copper surface which is horizontally aligned
inside a vacuum chamber (Fig. \ref{scheme}(a)) at a base pressure of $10^{-11}$\,mbar. We transport
ultracold clouds of $^{87}$Rb (T=\SI{1.5}{\micro\kelvin}) with optical tweezers to
a position \SI{200}{\micro\meter} above the surface and release the atomic cloud.
About $\sim10^6$ atoms are dropped in each experimental cycle onto the surface.
While the atomic cloud is falling towards the surface, we measure its EIT spectra. 

The ladder-type excitation scheme used for the EIT measurements \cite{Mohapatra2007, Kuebler2010}
is shown in Fig. \ref{scheme}(b). We probe the absorption on the $5S-5P$ transition with a weak
probe laser. The $5P$ state is strongly coupled to a highly excited $nS$ or $nD$
Rydberg state by means of a \SI{480}{\nano\meter} laser. While the frequency of the
probe laser is stabilized to the $5S-5P$ transition, the coupling laser can be
continuously scanned within a wide range of frequencies ($\pm$50 MHz). Whenever the
coupling laser is on resonance with a Rydberg state, the conditions for EIT are
satisfied and the atomic ensemble becomes transparent for the probe laser
\cite{Fleischhauer2005}. The presence of adsorbed atoms on the surface perturbs the
Rydbeg states and the resonance conditions for the coupling laser. We observe
pronounced energy shifts towards the surface which increase with the deposition of
atomic clouds. For the measurements we used $S$ and $D$ Rydberg states with principal
quantum numbers $n$ between 31 and 48.  

For our experiments we prepare clouds of $^{87}$Rb atoms in the $5S_{1/2}
(F\mathord=2,\,m_F\mathord=2)$ state in a setup described in \cite{Cano2011}. The
atomic cloud is loaded from a magneto-optical trap (MOT) into a Ioffe-Pritchard type
magnetic trap and cooled by forced radio-frequency evaporation to a
temperature of \SI{1.5}{\micro \kelvin}. The cloud with about $10^6$ atoms
is then loaded into an optical dipole potential of a focussed \SI{1064}{nm} laser
beam. By moving the focussing lens with an air bearing translation stage, the
optical tweezers transport the atoms from the preparation zone over a distance of
\SI{35}{mm} to a position above a copper surface. The tweezers are micropositioned
\SI{200}{\micro \meter} above the surface and the atomic cloud is released by
instantly ramping down the laser power. After 5\,ms of free fall and expansion, the
atomic cloud is imaged by absorption imaging. As the cloud falls, EIT spectra of
the atoms are taken. The imaging beam (probe beam for the EIT) has a Gaussian profile of 
\SI{7.5}{\milli\meter} FWHM and $\sim$\SI{100}{\micro\watt} power. The counter-propagating coupling beam is simultanouosly
focussed onto the cloud. It has an elliptical profile with $\sim$\SI{200}{\micro
\metre} and \SI{300}{\micro \metre} FWHM, respectively. The frequency stabilization
of the lasers with an absolute accuracy of better than \SI{\pm1.5}{\mega\hertz} is
described in \cite{Mack11}.

\begin{figure}
\centerline{\includegraphics[trim = 0 0 0 0, width=.5\textwidth]{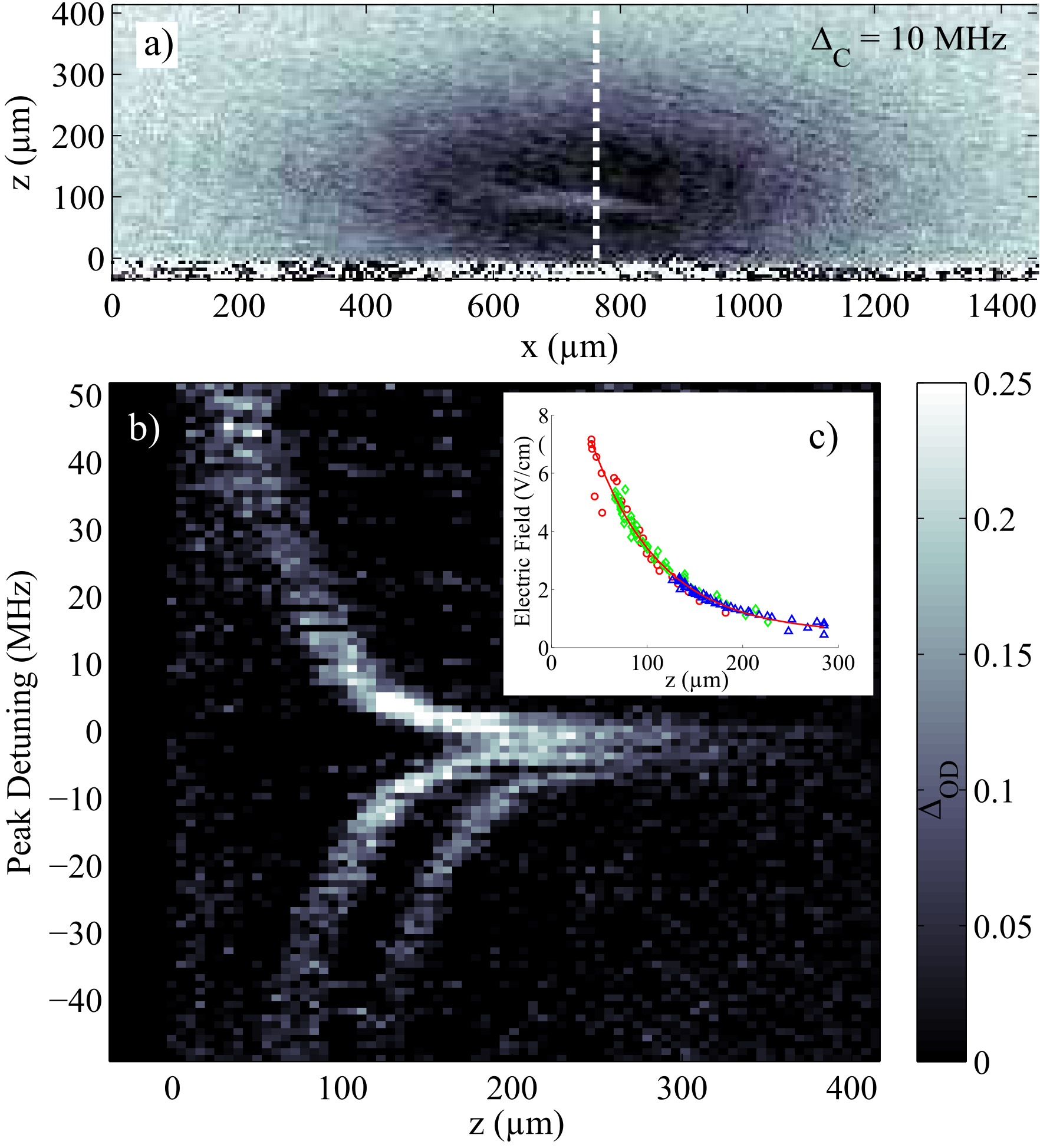}}
\caption{(a) Absorption image of an atomic cloud during free fall onto the surface
taken with the probe beam. The stronger the absorption, the darker the pixel in the
image. In this example, the coupling laser is blue detuned by 10 MHz from the $35D_{5/2}$ state of
$^{87}$Rb and produces a transparency window. The lateral extension of the window is
given here by the width of the coupling laser beam. For the data analysis we take
the $z$-position of the transparency window along the vertical dashed line. Similar
images have been taken for detunings ranging from \SI{-50}{\mega \hertz} to
\SI{+50}{\mega \hertz}, allowing the measurement of electrostatic fields as a
function of $z$. (b) Map of the relative optical density as a function of distance
to the surface (horizontal axis) and detuning of the coupling laser from the
unperturbed transition frequency  $5P_{3/2} (F\mathord=3,\,m_F\mathord=3)$ $
\rightarrow  35D_{5/2}$ (vertical axis). Each horizontal line shows a single
measurement. (c) Calculated electrostatic field as function of $z$ using the EIT
measurements on the $35D_{5/2}$ state. The different colors are the field strengths
obtained using the $|m_J| = 1/2$ (red circles), 3/2 (green diamonds), and 5/2 states
(blue triangles). The data of the three states follow the same curve, which is well
approximated by an exponential decay.}
\label{Map1}
\end{figure}

The measurements are illustrated in Fig. \ref{Map1}(a). The absorption image of an atomic
cloud during free fall shows a `transparency window' at the positions where the
coupling laser is resonant with the transition, in this case the $35D_{5/2} (m_J=
1/2)$ state. The image
shows that the resonance condition is satisfied only in a small window revealing a
spatial inhomogeneity of the energy shift. In this window the shift of the
Rydberg level equals the detuning $\Delta_C$ of the coupling laser, in this example
+10 MHz with respect to the unperturbed transition frequency. If the detuning of the
coupling laser is changed, the transparency window appears at a different distance
from the surface. 

We take a series of absorption images as a function of the detuning $\Delta_C$. The
laser frequency is varied in steps of \SI{1}{\mega\hertz} between consecutive
measurements. The results are summarized in Fig. \ref{Map1}(b), which shows 
the Stark shift of Rydberg states as a function of the distance $z$ to the surface.
Each horizontal line of Fig. \ref{Map1}(b) is obtained from the vertical column of absorption images, 
as indicated by the vertical dashed line in Fig. \ref{Map1}(a). Thus it shows the position of
the transparency window for different detunings. The three branches in Fig. \ref{Map1}(b)
correspond to the projections of the total angular momentum $J$ of the $35 D_{5/2}$ Rydberg state: $|m_J|=1/2, 3/2, 5/2$.

We now determine the electrostatic field above the copper surface using the measured
Stark shifts and comparing them with the theoretically calculated shift of Rydberg
levels in electrostatic fields. We calculate the Stark maps with the numerical
method of Ref. \cite{Zimmerman1979}. For our evaluation, we use an algorithm which
identifies the electrostatic fields that best fit the measured data. As shown in
Fig. \ref{Map1}(c), the results for the three different $|m_J|$ states lie on the same curve,
confirming the validity of our procedure. The shifts of the Rydberg states are thus
explained by static electric fields alone. The decay of the electrostatic field is
modelled here with an exponential function: $E(z) = E_0 \exp(-z/\sigma) +
E_\mathrm{res}$, where $z$ is the
distance to the surface, $\sigma$ is the decay length and $E_\mathrm{res}$ is a
residual, homogeneous electrostatic field that accounts for possible external field
sources. Figure \ref{Map1}(c)
shows the best-fit fields calculated with the three experimental curves of Fig.
\ref{Map1}(b). We repeated our measurements on $S$ and $D$ Rydberg states with principal
quantum numbers $n$ between 31 and 48 that reproduce the same behaviour. For
distances smaller than \SI{30}{\micro \meter}, the electrostatic field cannot be
determined reliably, as high field gradients over the size of one pixel of the
camera (\SI{5.6}{\micro \meter} in the object plane) lead to blurring of the
measured line shifts. 
We note that the electric field is also inhomogeneous along the $x$-axis. This is a result of the Gaussian distribution of the atomic clouds dropped onto the surface and of the residual roughness of the copper. In order to facilitate the evaluation of the changes of the field with time, all the measurements have been evaluated along the same line, as indicated in Fig. \ref{Map1}(a).

\section{Temporal evolution of the electrostatic fields of deposited adatoms}

\begin{figure}
\centerline{\includegraphics[ width=.5\textwidth]{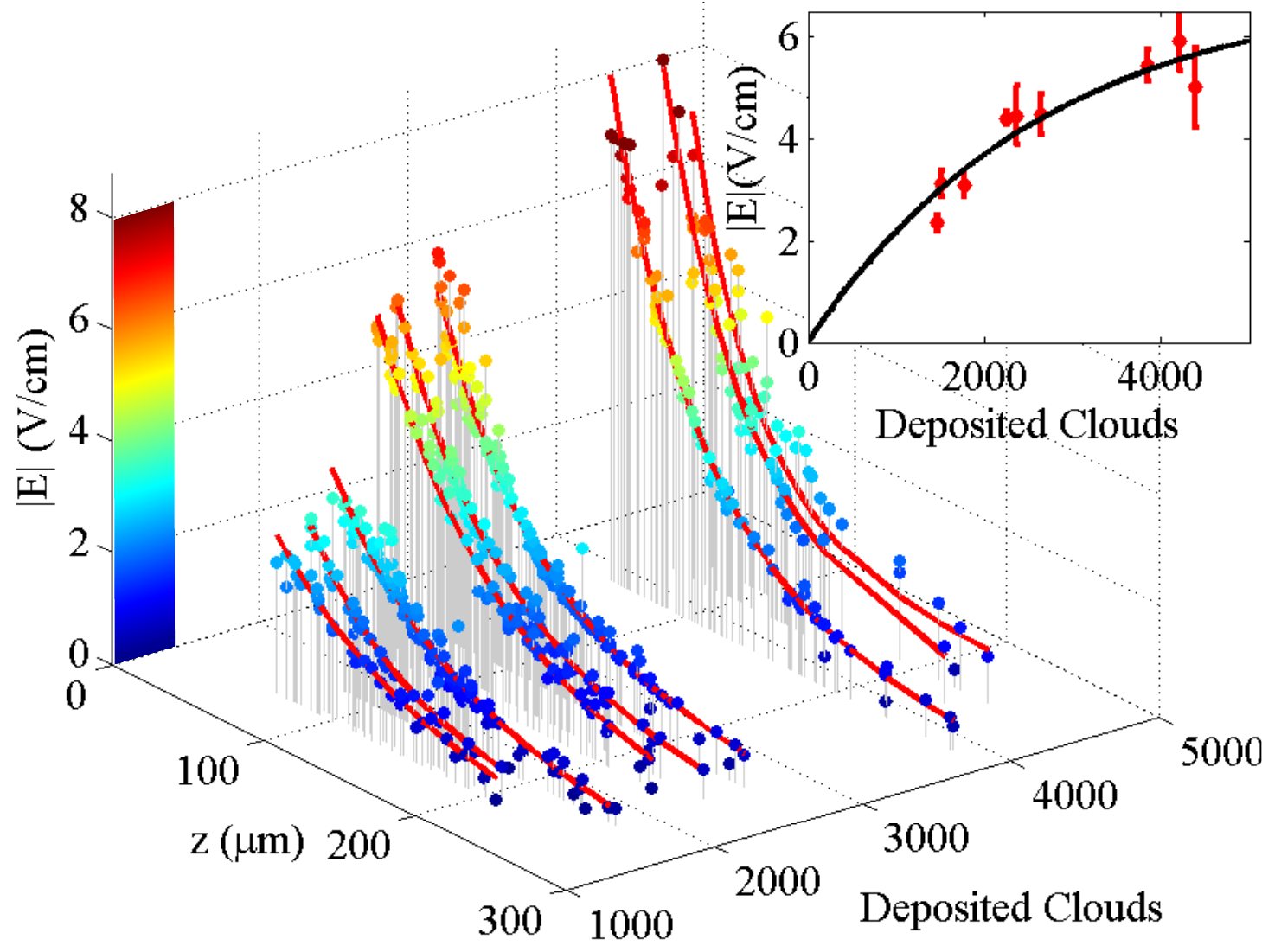}}
\caption{Measured electrostatic field as a function of $z$ and the number of
experimental cycles carried out. Inset: Measured electrostatic field at a distance
of \SI{80}{\micro\meter} from the surface. We observe an increase of the electric
field due to adsorption of Rubidium onto the copper surface. The saturation builds
up after the deposition of few thousand atomic clouds.}
\label{E_vs_time}
\end{figure}

We evaluate the evolution of the electrostatic field close to the surface as atom
clouds are repeatedly deposited on it. Figure \ref{E_vs_time} summarizes the results. The diagram
shows the electric field as a function of the distance from the surface and the
number of deposited atomic clouds. Different colors correspond to different strengths of the electric field. The red lines
are exponential fits that we use for determining the electric field as in Sec. II.
The inset shows the increase of the measured electric field with the number of deposited
atoms for a distance of \SI{80}{\micro\meter} from the surface. The magnitude of
the field increases with the number of experimental runs. However, we observe a
saturation after about ten days of experiments during which we released
approximately $5\cdot 10^9 $ atoms on the copper surface. Based on our measurements,
we estimate that it takes as little as few hundred experimental runs to produce an
electric field of \SI{1}{\volt/\centi\meter} at a distance of
\SI{30}{\micro\meter}, assuming a zero-field at the beginning of the experiments.
This field produces a level shift of \SI{-2}{\mega \hertz} on the $35S_{1/2}$ and of
\SI{-49}{\mega \hertz} on the $55S_{1/2}$ state. This is much larger than the
linewidth of a high-Q stripline resonator, making cavity-QED experiments problematic
by shifting the atoms out of the cavity resonance.

\section{Conclusion}  
Our measurements show that neutral atoms adsorbed on a metal surface cause
electrostatic fields on the order of \SI{1}{\volt/\centi\meter} after as little as a
hundred repetitions of a cold atom experimental cycle. Adsorbate fields have also been 
observed on dielectric surfaces\cite{Abel2011}. This sets serious limitations
on the feasibility of cavity QED experiments with Rydberg atoms and coplanar
cavities. Also dispersion forces between Rydberg atoms and planar surfaces
\cite{Crosse2010a} are masked by the strong electric fields of adsorbates.   A search
for strategies to correct for this problem is therefore very important for atom
chips. A possible solution could be the cleaning of the surface whenever the
electrostatic fields due to adsorbates become harmful. For example, regular heating
of the surface cause adsorbed atoms to diffuse.
Another possibility would be photodissorption of the adsorbed atoms, but given the 
work function of metals, this would require light in the far ultraviolet range.
Given the fast appearance of detrimental adsorbate fields, an open question is still 
if there are cleaning techniques which can be applied quickly between experimental 
cycles. 
A workaround for this problem would be the development of experimental techniques that 
avoid deposition of atoms onto the surface or using surface coatings with materials on which 
no adsorbate fields have been observed \cite{Kuebler2010}.
While atoms on surfaces have undesired effects on cold atom experiments, it is worth mentioning that adatoms may be useful to control electric properties of surface layers. For example, alkali metal adsorbates have been used to engineer the electronic structure of graphene \cite{Ohta2006, Jin2010}.
  
Rydberg EIT can be used for a sensitive measurement of electric fields. In
combination with micropositioning of atomic clouds by optical tweezers or magnetic
conveyor belts in a scanning probe configuration \cite{Gierling2011} three dimensional imaging of
the electric field distribution is feasible.  However, the measurement technique contaminates the surface, which must be taken into account.

\section*{Acknowledgements}
The authors would like to thank Thomas Judd for useful discussions.
This work was supported by the European Research Council (Socathes) and the Deutsche
Forschungsgemeinschaft (SFB TRR21). The authors acknowledge additional support from
the ev. Studienstiftung Villigst e.V., and the Baden-W\"urttemberg-Stiftung through
the ``Kompetenznetz Funktionelle Nanostrukturen''.

\end{document}